\documentclass[conference]{IEEEtran}
\IEEEoverridecommandlockouts
\usepackage{cite}
\usepackage{amsmath,amssymb,amsfonts}
\usepackage{algorithmic}
\usepackage{graphicx}
\usepackage{textcomp}
\usepackage{xcolor}
\usepackage{todonotes}
\usepackage{subcaption}

\newcommand{\ffig}[1]{Figure~\ref{#1}}

\def\BibTeX{{\rm B\kern-.05em{\sc i\kern-.025em b}\kern-.08em
    T\kern-.1667em\lower.7ex\hbox{E}\kern-.125emX}}
\begin{document}

\title{Integrating embedded neural networks and self-mixing interferometry for smart sensors design

\thanks{This project has received financial support from the CNRS through the MITI interdisciplinary programs}
}

\author{
\IEEEauthorblockN{1\textsuperscript{st} Pierre-Emmanuel Novac}
\IEEEauthorblockA{
\textit{Université Côte d'Azur, LEAT}\\
Nice, France \\
pierre-emmanuel.novac@univ-cotedazur.fr}
\and
\IEEEauthorblockN{2\textsuperscript{nd} Laurent Rodriguez}
\IEEEauthorblockA{
\textit{Université Côte d'Azur, LEAT}\\
Nice, France \\
laurent.rodriguez@univ-cotedazur.fr}
\and
\IEEEauthorblockN{3\textsuperscript{rd} Stéphane Barland}
\IEEEauthorblockA{
\textit{INPHYNI, CNRS, Université Côte d'Azur}\\ %
Nice, France \\
stephane.barland@univ-cotedazur.fr}

}

\maketitle

\begin{abstract}
  Self-mixing interferometry is a measurement approach in which a laser beam is re-injected into the emitting laser itself after reflection on a target.
  Information about the position of the target can be obtained from monitoring the voltage across the laser.
  However, analyzing this signal is difficult.
  In previous works, neural networks have been used with great success to process this data.
  In this article, we present the first prototype of an integrated sensor based on self-mixing interferometry with embedded neural networks.
  It consists of a semiconductor laser (acting both as light emitter and detector) equipped with an embedded platform for data processing.
  The platform includes an ADC (Analog-to-Digital Converter) and an STM32L476RG microcontroller.
  The microcontroller runs the neural network in charge of reconstructing the displacement of a target from the interferometric signal entering the ADC.
  We assess the robustness of the neural network to unwanted signal amplitude variations and the impact of different network weights quantization choices required to run the network on the microcontroller.
  Finally, we provide a demonstration of target displacement reconstruction fully running on the embedded platform.
  Our results pave the way towards robust, low power and versatile sensors based on self-mixing interferometry and embedded neural networks.

\end{abstract}


\section{Introduction}

Self-mixing interferometry is a well-established and versatile measurement approach in which a laser beam is re-injected into the emitting laser itself after reflection on a target \cite{giuliani2002laser,kane2005unlocking,donati2012developing,taimre2016laser,li2017laser,rakic2019sensing}.
Thus, optical interferences between the emitted and reflected beams take place inside the laser itself, altering its operation point.
In normal operating conditions, a semiconductor laser emits a monochromatic beam whose intensity (and all other properties) are fully determined by the laser construction parameters and by the amount of energy brought into the laser, \textit{ie} the electrical pumping current.
In the case of self-mixing interferometry, the interference condition between the emitted and the re-injected beam constitutes an additional parameter which influences the intensity emitted by the laser.
Thus, information about the position of the target (or many other properties, see \textit{eg} \cite{brambilla2020versatile} for an overview) can be obtained from monitoring the laser operation point, by recording the voltage across the laser diode as a function of time.
Recovering the displacement of the target per time unit is in principle a simple problem.
If the displacement is less than an optical wavelength a physical model can be fit to the data and the displacement can be recovered.
Alternatively, if the displacement is of many wavelengths, counting minima and maxima of interference can lead to reconstruction of the displacement.
In practice though, this signal analysis is often complex and remains error prone due to detection noise, very large signal bandwidth (even for very simple displacement patterns) or variations of the reflectivity of non-cooperative targets.
Many hardware or software approaches have been designed to mitigate these long-standing issues \cite{norgia2001interferometric,zabit2010adaptive,bernal2014robust,arriaga2014speckle,usman2019detection,siddiqui2017all,usman2019detection,bernal2021toward}.
As was shown recently \cite{barland2021convolutional} a relatively small convolutional neural network can recover the displacement of the target from the voltage across the laser diode with very high precision.
This method has recently been extended to a multichannel measurement scheme, which results in much improved measurement availability \cite{matha2023high}.
In \cite{ahmed2019self} neural networks were used for signal pre-processing, in \cite{kou2020fringe} neural networks were used for fringe discrimination in the signal and \cite{an2022measuring} discussed the use of neural networks for parametric conditions estimation.

Further development of the self-mixing approach as a general purpose sensor \cite{brambilla2020versatile} suitable for standalone operation without involvement of specifically trained personnel therefore requires bringing suitable neural networks to a widely available microcontroller platform.

Such a device is the STMicroelectronics STM32L476RG, a very popular low-power 32-bit microcontroller which can be used to run the inference of tiny deep neural networks.
To achieve this, a software framework that enables the deployment of deep neural networks on microcontrollers must be used.
Popular frameworks include TensorFlow Lite for Microcontrollers and STM32Cube.AI, but we chose to use Qualia-CodeGen\cite{qualia} for its low memory footprint and its ease of customization\cite{microaiarticle}.
The neural network is quantized to fixed-point numbers in order to keep the memory footprint low and to obtain the best performance.

Furthermore, the embedded platform can be designed so that it captures the signal directly through an ADC (Analog-to-Digital Converter), rather than requiring an external capture device to acquire the data on a workstation.
Finally, the use of a microcontroller dedicated to this task, with a high level of control over the data capture, paves the way to real-time operation.
This enables the development of a first prototype of a self-contained device for the sensing of small displacements through self-mixing interferometry.

Our contributions are as follows:
\begin{itemize}
  \item design of a tiny deep neural network for small displacement measurements through self-mixing interferometry,
  \item evaluation of fixed-point quantization of the deep neural network,
  \item evaluation of embedded performances after deployment on microcontroller,
  \item design of a first prototype of a device for small displacement sensing.
\end{itemize}

The rest of this article is organized as follows.
Section \ref{sec:method} explains the deep neural network, its quantization and deployment on the microcontroller.
Section \ref{sec:setup} describes the optoelectronics setup and the embedded platform.
Section \ref{sec:results} presents results on the prediction performance of the neural network before and after quantization, as well as results on the embedded execution.
Finally, section \ref{sec:conclusion} concludes our work and provides future perspectives.

\section{Proposed method}
\label{sec:method}

\subsection{CNN based reconstruction}
\begin{figure}[ht]
    \center
    \includegraphics[width=0.8\linewidth]{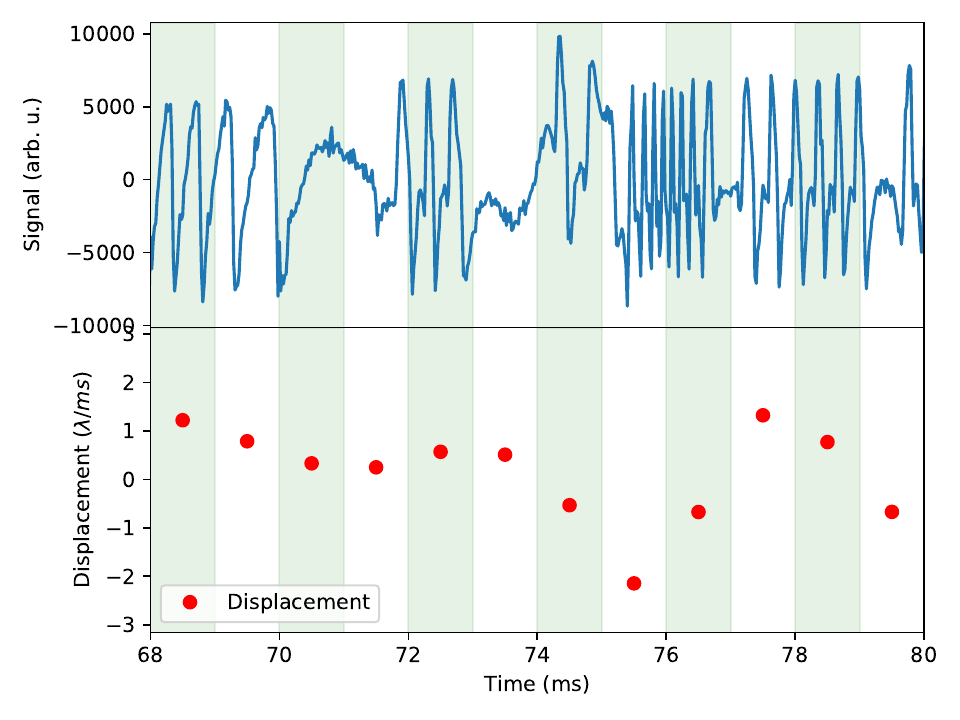}
	\caption{Example of the regression task to be realized. Top: the interferometric signal is measured as a voltage at the edges of a semiconductor laser diode in the course of time. Bottom, the displacement of the target during the same time frame. The task is to reconstruct the displacement of the target (red dots) from the signal during the corresponding time segment (the alternating light green and white stripes)}    
    \label{fig:example}
\end{figure}

The task to be realized by the network is represented on \ffig{fig:example}. On the basis of segments of the interferometric signal in the course of time, the network must infer the value of the displacement of the target during the considered time interval. One of the peculiarities of this signal is that, since it consists essentially of interference fringes taking place in a nonlinear medium, the amplitude of the signal is not significant in terms of the displacement of the target but only in terms of its reflectivity. Thus, the information which must be extracted from the signal resides in the shape of the fringes, their more or less peaked aspect, their orientation and their repetition rate.

The network we start from was originally described in \cite{barland2021convolutional} and consists of a an alternating stack of four convolutional and four maxpooling layers followed by two fully connected layers for the final regression. In the original work, the network (featuring $5.7*10^4$ parameters) processed segments of 256 points acquired at a sampling rate of 250kS/s. This network was trained and evaluated on many segments of interferometric signal (data available at \cite{barland2022displacement} with details in \cite{barland2021convolutional}) of comparable amplitude, the whole data set being scaled to its standard deviation.

In order to run the network on the STM32 platform, we reduced it so that it operates on non-overlapping windows of 48 samples acquired at a rate of 48kHz. In addition, a batch normalization layer has been added between each convolution layer (kernel size of 3) and the following max pooling layer (pool size of 2). ReLU activation functions are used everywhere except for the final regression layer which is linear. The resulting network features 21601 tunable weights.

\subsection{Network training}

As we mentioned above \ffig{fig:example}, the peak to peak amplitude of the interferometric signal does not contain information about the displacement of the target and may instead be altered by other factors (reflectivity of the target or alignment conditions). Thus, the network must be trained to be robust to variations of the amplitude of the signal. To optimize the network for the reconstruction task at hand, we start from the data set available in \cite{barland2022displacement} which consists of 195011 segments of interferometric signal with their corresponding true displacement value. This data set was downsampled so that each segment of signal consists of 48 points instead of 256. In addition, the original data set is normalized by its standard deviation. Here, to simulate spurious signal amplitude variations, we apply during training a random multiplication factor comprised between 6 and 60 to each segment of input signal. The network is trained during 40 epochs with the RMPSprop optimizer and an initial learning rate of $10^{-3}$ which is decreased progressively down to $6*10^{-5}$ on learning plateaus, 20\% of the data set being kept for validation.

\subsection{Artificial neural network quantization}

After training the artificial neural network (ANN), its weights and activations are quantized. The inputs are also quantized.
This enables a reduction both in terms of memory and latency.
Indeed, going from 32-bit floating-point numbers to 16-bit fixed-point numbers divides the memory usage of the weights (in ROM), the activations (in RAM) and the inputs (in RAM) by 2.
Furthermore, computation with fixed-point numbers can be better optimized to be performed faster than floating-point numbers.

The quantization of the neural network follows the method presented in \cite{microaiarticle}.
Post-training quantization with floor rounding mode, a power-of-two scale factor and a symmetric range is used. The same scale factor is applied to weights, biases, activations and inputs.
For 16-bit quantization, the empirically chosen format is Q7.9, and for 8-bit quantization the chosen format is Q2.6.
Qx.y means x bits for the integer part (including sign) and y bits for the fractional part.

\subsection{Deployment on microcontroller}

The Qualia-CodeGen tool from the Qualia framework\cite{qualia} is used in order to generate a portable C library for inference from the trained Keras model.
Qualia-CodeGen generates an inference function for each layer of the neural network, along with the layers call chain.
In order to speed the inference up, the CMSIS-NN\cite{CMSISNN} library is used.
CMSIS-NN contains optimizations for computation of neural network operations such as convolutions, and makes use of specific instructions when available
(e.g. SMLAD to perform two 16-bit integer multiply-accumulate operations in a single cycle).
The generated C code is then compiled with the GCC compiler, using the \texttt{-Ofast} optimization level, and uploaded to the microcontroller.

\section{Experimental setup}
\label{sec:setup}

\begin{figure}[ht]
  \includegraphics[width=\linewidth]{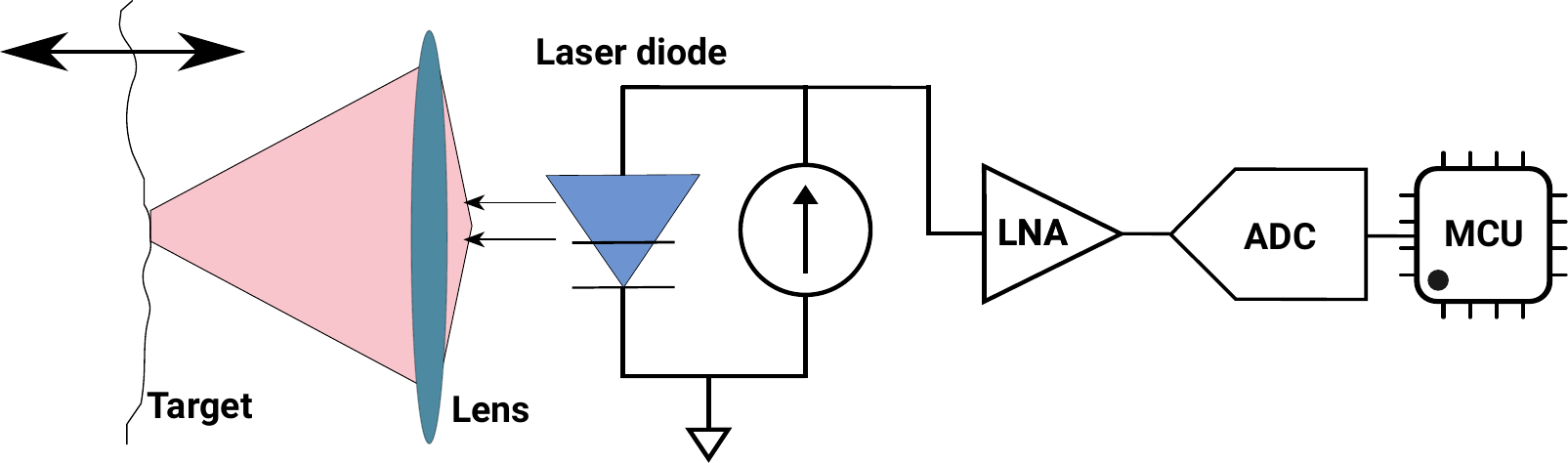}
  \caption{Experimental setup, LNA: Low-Noise Amplifier, ADC: Analog-to-Digital Converter, MCU: MicroController Unit}
  \label{fig:setup}
\end{figure}

\subsection{Optoelectronics setup}

The experimental arrangement is shown schematically on \ffig{fig:setup}. From the optical point of view it consists of a semiconductor laser driven slightly above its coherent emission threshold by a stabilized current source emitting a beam of about 1~$mW$ power at a wavelength $\lambda=1.310~\mu m$. The beam is focused by a lens on a target at a distance of about 15~cm and a tiny part of it (of the order of $10^{-4}$) is re-injected back into the laser. In this case the target is a simple speaker whose displacement has been calibrated so that the voltage at the edges of the speaker can be used as a proxy for the displacement of the target. The details of this optical system are available in \cite{barland2021convolutional} and we omit them here for brevity. The interferometric signal is obtained by measuring the voltage at the edges of the laser diode. This voltage consists of a large (about 1~V) continuous component superimposed with much smaller fluctuations (sub-mV) which contain the relevant part of the signal. We amplify the voltage with an AC-coupled (approximately $10^4$ gain) amplifier and the resulting signal is sent to the embedded platform.

\subsection{Embedded platform}

\begin{figure}[ht]
    \center
    \includegraphics[width=0.7\linewidth]{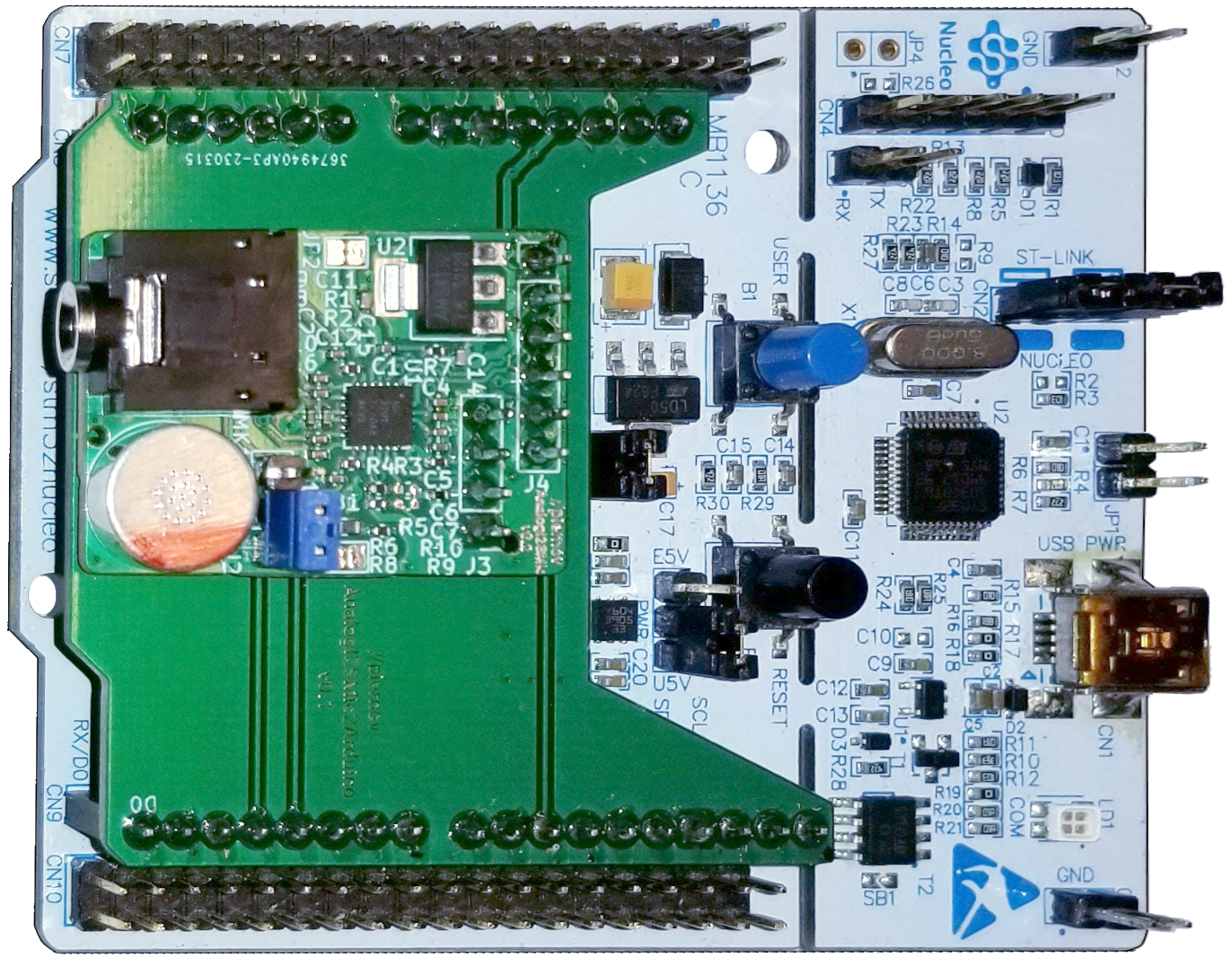}
    \caption{Embedded platform}    
    \label{fig:embedded_platform}
\end{figure}

Our prototyping embedded platform consists of a Nucleo-L476RG board (white PCB) mounted with a custom capture interface (green PCB), as seen in Fig. \ref{fig:embedded_platform}.
The Nucleo-L476RG board is designed around an STM32L476RG microcontroller with a Cortex-M4F core running at 80~MHz, 1~MiB of Flash memory and 128~KiB of SRAM.

The custom capture interface is originally designed for audio applications with microphones.
However, it can also be used to capture other AC signals, although the ADC only supports standard audio sampling rates.
The TLV320ADC3101, which contains ADCs, amplification stages and filters, captures the input signal and sends it to the microcontroller through an I$^2$S interface.
The TLV320ADC3101 is configured with a 10~dB analog gain to the input signal, a sampling rate of 48~kHz, and a resolution of 16 bits.
Only one channel is used.

The signal is fed through a 3.5~mm jack connector.
An external low-noise amplifier with a gain of approximately $10^4$ is placed between the laser and the capture interface input jack.
The amplitude of the signal at the output of the low-noise amplifier varies depending on the calibration of the optoelectronics setup, however a good calibration provides around 400~mV peak-to-peak.

After being captured by the ADC, the signal is divided by $2^5$ in software as an initial guess to better match the amplitude of the signal to the ANN input layer. No z-score normalization is applied.

For evaluation of the embedded platform using data from the test dataset, the output of the low-noise amplifier is replaced with the audio output of a laptop which streams the interferometric signal from the test dataset.

\section{Results}
\label{sec:results}

\subsection{Robustness to amplitude variations on workstation}

\begin{figure}[htbp]
\centerline{\includegraphics[width=0.5\textwidth]{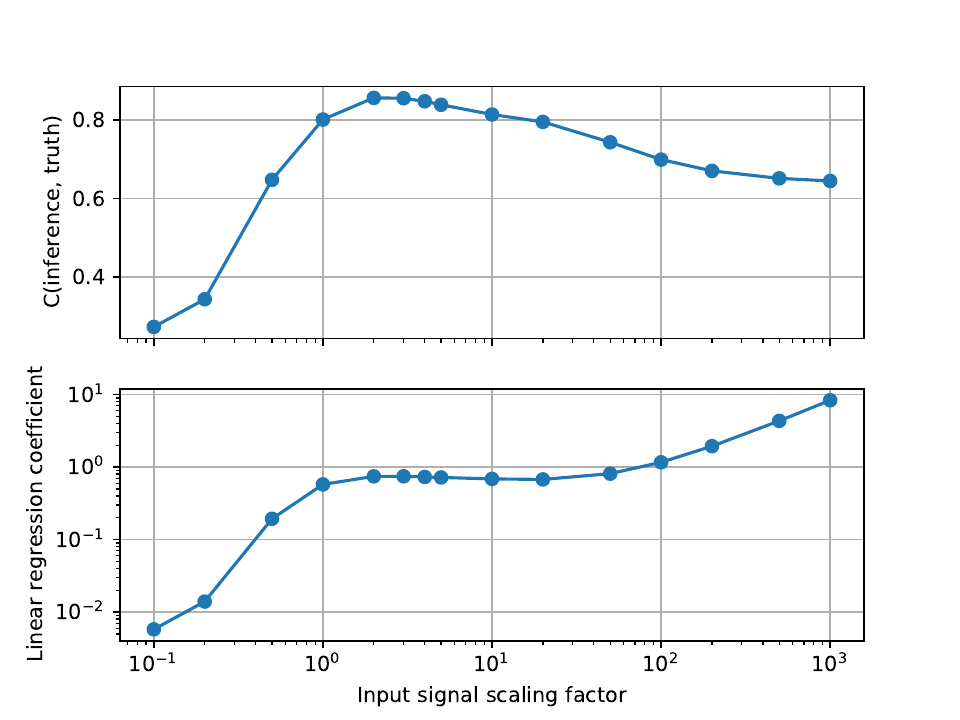}}
	\caption{Scaling signal influences the inference quality.}
	\label{fig:scaling}
\end{figure}

As outlined in \ref{sec:method}, the network must be able to reconstruct the displacement of the target even in presence of unwanted variations of the signal amplitude. We assess the performance of the network on workstation by comparing the true and the inferred displacement on a validation data set (never used during training) for different magnification factors applied to the input signal. The results are shown on \ffig{fig:scaling}. One good metric for performance is of course the Pearson correlation coefficient between the true and inferred displacement, shown on top of \ffig{fig:scaling}. The correlation is found to be above 0.8 for scaling factors between 1 and 20, with a maximum of about 0.85 for amplification factors between 2 and 3. These values can be compared to an optimal Pearson coefficient of 0.9 with a much larger network in \cite{barland2021convolutional}. 

Beyond correlation, the performance of the network can be assessed by measuring the linear regression coefficient between the inferred and the true displacement, as shown on the bottom panel of \ffig{fig:scaling}. In this case, the optimal operation range is found for amplification factors between 2 and 100, whereas the correlation decreases for much smaller amplifications indicating a noisier inference, still keeping a good average precision. In the optimal range, the regression coefficient is found to be slightly smaller than 1, which would be the most precise inference. This is related to scarcity of training data for the larger absolute values of displacement which imply a saturation of the inference in these regions.

The existence of plateaus in both the correlation and the regression coefficients indicates that the network is robust to about one order of magnitude of amplitude variations.

In the results that follow, the input data is multiplied by 3 in order to use the optimal range.

\subsection{Evaluation of quantization}

The effect of quantization of the neural network is evaluated by generating the C code with Qualia-CodeGen.
Three configurations are evaluated: floating-point numbers with no quantization (float32), fixed-point 16-bit quantization with format Q7.9 (int16), and fixed-point 8-bit quantization with format Q2.6.
The inference is then performed on a workstation with the test dataset.

\begin{figure}[ht]
    \includegraphics[width=0.9\linewidth]{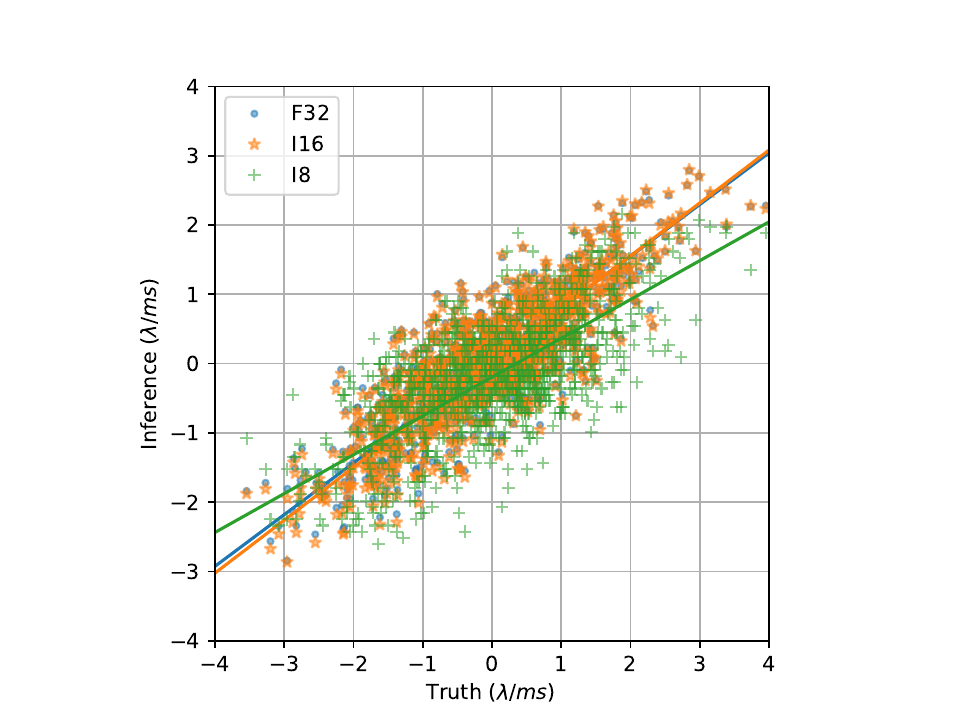}
	\caption{Predictions for float32, int16 and int8 models vs. ground truth. The solid lines are optimal linear fits with coefficients 0.74, 0.76 and 0.56 for float32, int16 and int8 respectively. }
    \label{fig:quantization_dispersion}
\end{figure}

\begin{table}[ht]
  \centering
  \caption{Pearson Correlation Coefficient for float32, int16 and int8 predictions compared to ground truth}
    \begin{tabular}{|c|c|c|r|r|r|}
      \hline
      \textbf{Data format} & \textbf{Pearson Correlation Coefficient}\\
      \hline
      float32 & 0.8551\\ %
      \hline
      int16 Q7.9 & 0.8556\\ %
      \hline
      int8 Q2.6 & 0.7076\\ %
      \hline
    \end{tabular}
  \label{tab:quantization_correlation}
\end{table}

Fig. \ref{fig:quantization_dispersion} shows the scatter plot for predictions of the neural network for int8 quantization (green), int16 quantization (orange), and non-quantized float32 (blue) versus the ground truth.
Table \ref{tab:quantization_correlation} provides the Pearson correlation coefficient with respect to the ground truth for each configuration.

The difference in correlation coefficient between float32 and int16 is only $+0.0005$, we consider this not to be significant.
The scatter plot shows a slight shift in the distribution towards lower values from float32 to int16.
This can be explained by the floor rounding mode used for quantization.
A less biased result may be obtained by using round to nearest instead.

On the other hand, int8 quantization causes a significant loss in prediction performance as the correlation coefficient drops by 0.1474 compared to float32.
This highlights the limits of such a simple quantization scheme.
In order to improve the results of 8-bit quantization, the scale factor must be chosen separately for each layer, and it must also be decoupled for the inputs, the weights and the activations.
Quantization-aware training could also be considered.

\subsection{Evaluation of embedded constraints}

\begin{table}[ht]
  \centering
  \caption{Latency and memory consumption on STM32L476RG for float32, int16 and int8 inference, with and without batch normalization, and with and without CMSIS-NN}
  \resizebox{\linewidth}{!}{
    \begin{tabular}{|c|c|c|r|r|r|}
      \hline
      \textbf{Data format} & \textbf{BatchNorm} & \textbf{CMSIS-NN} & \textbf{Latency} & \textbf{ROM} & \textbf{RAM}\\
                         &                      &                   & \textbf{(ms)}         & \textbf{(kB)} & \textbf{(kB)}\\
      \hline
      float32 & No & No & 14.8 & 136.672 & 12.120\\ %
      \hline
      int16 & No & Yes & 6.2 & 95.840 & 11.336\\ %
      \hline
      int16 & No & No & 13.2 & 94.488 & 9.816\\ %
      \hline
      float32 & Yes & No & 15.2 & 138.776 & 13.592\\ %
      \hline
      int16 & Yes & Yes & 6.8 & 97.104 & 12.072\\ %
      \hline
      int16 & Yes & No & 13.9 & 95.752 & 10.552\\ %
      \hline
      int8 & Yes & Yes & 7.4 & 78.456 & 10.200\\ %
      \hline
      int8 & Yes & No & 14.2 & 74.184 & 8.680\\ %
      \hline
    \end{tabular}
  }
  \label{tab:embedded_measurements}
\end{table}

Table \ref{tab:embedded_measurements} shows latency and memory for different configurations of the neural network when deployed on the microcontroller.
The latency is measured as the time taken to perform the inference, after collecting a window of data in the input buffer.
For the memory footprint, both ROM and RAM footprints are provided. They are extracted from the memory allocation in the firmware after compilation.
The ROM contains the program instructions and the neural network weights.
The RAM contains the input buffer and each layer's output buffer, all allocated statically.
When possible, i.e., on independent layers, buffers share a common memory space.
Note that some memory is also used for other parts of the firmware, e.g., I2S and serial communications.

In these results, we consider cases with and without batch normalization layers.
For now, the batch normalization layers are not fused with the preceding convolutional layer.
As can be seen, the case with batch normalization is slower by a few tenths of a millisecond, and has a slightly higher memory footprint.
However, in the future, we can assume that the inference time with fused batch normalization will be the same as without batch normalization.
Furthermore, this will have the same effect on the memory footprint.

We also compare the effect of fixed-point quantization, with and without CMSIS-NN optimizations.
As expected, 16-bit quantization significantly reduces the memory footprint compared to 32-bit floating point by approximately 30\% for the ROM.
Without CMSIS-NN, the reduction in inference time is of less than 10\%, however using CMSIS-NN allows for a decrease of 55\%.

8-bit goes further in terms of memory footprint reduction, by approximately 43\% when compared to 32-bit floating point.
Unfortunately, the inference time is not reduced in this case, we suspect that the packing and unpacking of data may not be as optimized.

All of these configurations fit well within the ROM (MCU limit: 384~kiB) and RAM (MCU limit: 128~kiB) constraints, however the latency is well above our real-time constraint of 1~ms.
A more optimized neural network and a higher-performance microcontroller may help reaching this target.

Still, the best performing configuration is using 16-bit quantization with CMSIS-NN optimization.
Furthermore, we already concluded in the previous section that 16-bit quantization does not significantly deteriorate the quality of the predictions over 32-bit floating point.

\subsection{Live inference}

\begin{figure}[htbp]
  \begin{subfigure}[t]{\linewidth}
    \includegraphics[width=\linewidth]{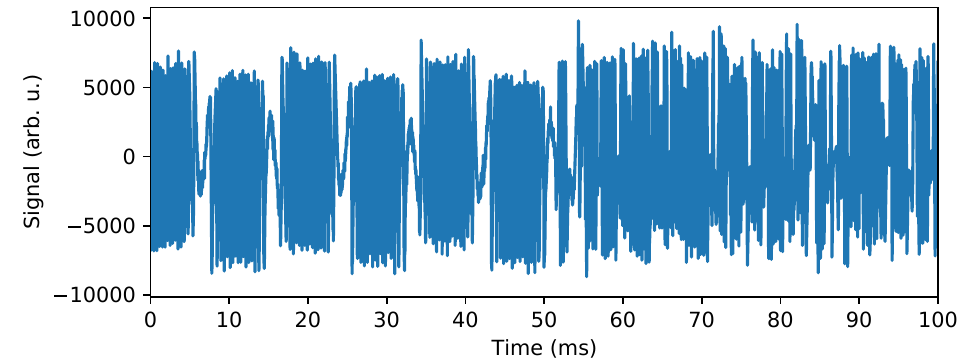}
    \caption{Interferometry signal}
    \label{fig:inference_signal}
  \end{subfigure}
  \begin{subfigure}[t]{\linewidth}
    \includegraphics[width=\linewidth]{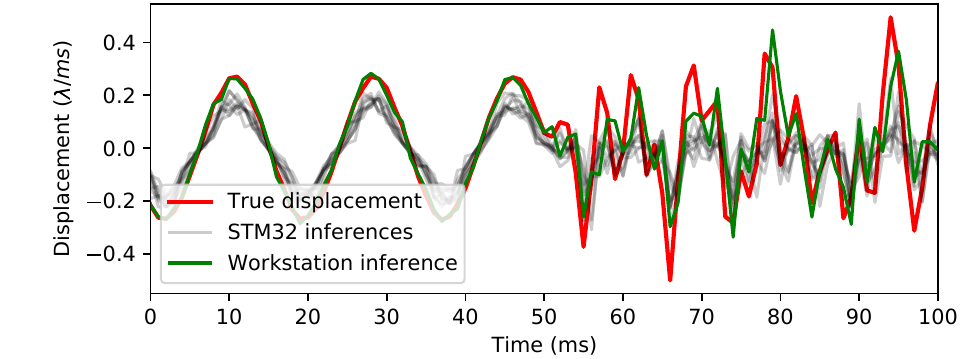}
    \caption{Predictions and ground truth}
    \label{fig:inference_preds}
  \end{subfigure}
	\caption{Example of live inference on workstation (red) and embedded platform (gray) to be compared to ground truth (red).}
\end{figure}

To demonstrate the experimental setup, we selected a displacement signal composed of a sinusoidal signal for the first half and a more chaotic signal for the second half (in red in Fig. \ref{fig:inference_preds}).
The associated interferometry signal that the deep neural network takes as inputs can be seen in Fig. \ref{fig:inference_signal}.
On the workstation, the signal is interpreted as is by the neural network (in floating point representation) and the predictions are shown in green in Fig. \ref{fig:inference_preds}.
On the embedded platform (STM32), the signal goes through the digital-to-analog converter of the workstation, is then transmitted as an analog signal through an audio cable to the embedded platform's capture interface.
The capture interface amplifies the signal, samples it and performs analog-to-digital conversion, for the microcontroller to handle the inference over the digitized signal.
This introduces variability in the signal, and therefore in the predictions. Fig. \ref{fig:inference_preds} shows several of these inferences in gray.

We can see that the predictions do not match the ground truth exactly, but the reconstruction is still visually similar.
In the sinusoidal part, the reconstruction on the workstation (in green) follows the ground truth (in red) very closely.
The predictions on the microcontroller (in grey) have a similar shape, but they are scaled down a little bit.
We suspect this is due to the fact that the signal at the input of the neural network on the microcontroller does not have the exact same amplitude as on the workstation, due to the multiple transformations and amplification of the signal.
For the chaotic part, the reconstruction looks a bit less precise but the shape stays similar.

\begin{figure}[htbp]
\centerline{\includegraphics[width=0.5\textwidth]{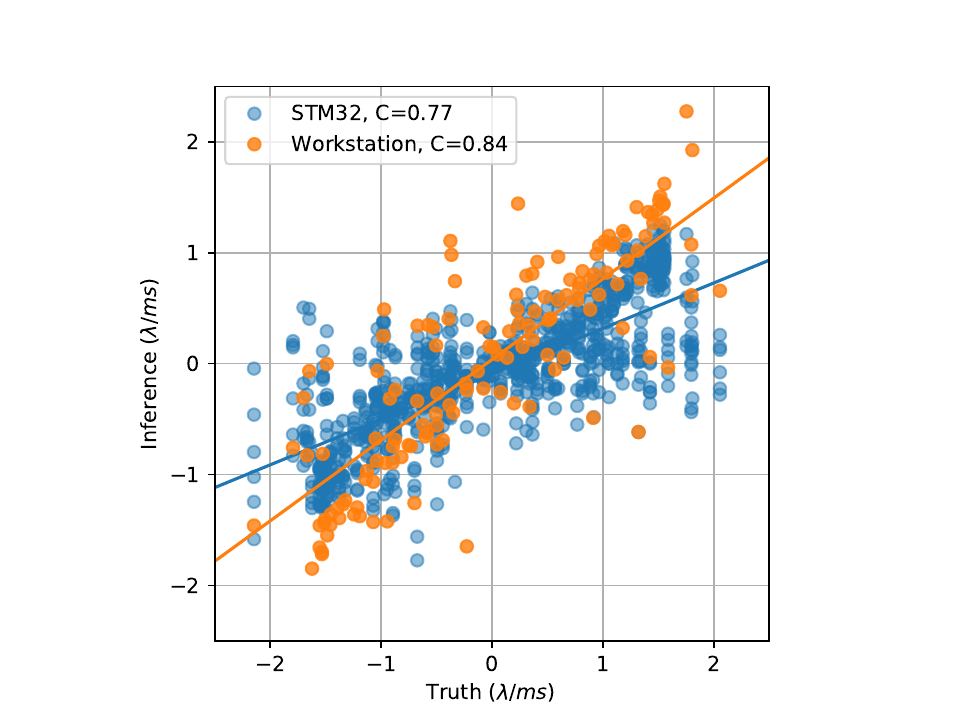}}
	\caption{Comparison of correlations between the true displacement and the displacement inferred on workstation and on the STM32 platform. The solid lines are optimal linear fits with coefficients 0.84 on the workstation and 0.41 on the STM32.}
\label{fig:correlations}
\end{figure}

Fig. \ref{fig:correlations} shows the scatter plot of the predictions vs. the ground truth, both for the predictions from the workstation and from the embedded platform.
Overall, the predictions from the workstation are of better quality than from the embedded platform.
The workstation (using floating point representation) provides predictions for this signal with a 0.84 correlation compared to the ground truth, while for the embedded platform the correlation with the ground truth is of 0.71.
We suspect that the lack of control over the exact amplitude of the signal and the lack of control of the exact sampling time, as well as the added noise due to the analog transmission can cause these differences.

\section{Conclusion}
\label{sec:conclusion}

In this article, we presented a first prototype of an embedded device for small displacement measurement through self-mixing interferometry.
To achieve this, a deep neural network compatible with the embedded constraints was designed and evaluated.
Then, the neural network was quantized in order to reduce the memory footprint and inference latency.
Finally, the neural network was deployed on an STM32L476RG microcontroller and evaluated in terms of memory footprint and latency.
Results showed that the signal reconstruction could achieve a 0.8556 correlation with the ground truth and that the memory footprint is kept below 10\% of the microcontroller capabilities.
However, real-time constraints are not met yet due to the inference latency being too high.

Future works will first focus on striving for real-time computation, i.e., reduce the latency below 1~ms.
This will include optimizations of the neural network to reduce the amount of computation, and evaluation of more powerful microcontrollers.
If the real-time constraint still cannot be met, a hardware accelerator could be used instead.
Furthermore, the issue related to the robustness of the network to variations of amplitude must be solved.
Apart from trying to make the network more robust, it may also be possible to leverage the automatic gain control of the TLV320ADC3101.
Finally, the embedded platform currently does not provide any synchronization mechanism.
An external trigger will be implemented to trigger an interrupt in the microcontroller and synchronize the start of the measurement.

Beyond that, the present work opens the way to many possible applications of AI-enabled self-mixing sensors including flow cytometry, microfluidics or environmental conditions classification.

\bibliographystyle{IEEEtran}
\bibliography{bibliography}

\end{document}